\documentclass[a4paper]{jpconf}
\usepackage{graphicx}
\usepackage{cite}
\usepackage{xspace}

\newcommand{\meg}{$\mu \rightarrow e\gamma$\xspace}
\newcommand{\mte}{$\mu \rightarrow eee$\xspace}

\begin{document}
\title{A novel experiment searching for the lepton flavour violating decay $\mu \rightarrow eee$}

\author{Niklaus Berger}

\address{Physics Institute, University of Heidelberg, Philosophenweg 12, 69121 Heidelberg, Germany}

\ead{nberger@physi.uni-heidelberg.de}

\begin{abstract}
Since the discovery of neutrino oscillations it is known that lepton flavour is not conserved. Lepton flavour violating processes in the charged lepton sector have so far however eluded detection; as they are heavily suppressed in the standard model of particle physics, an observation would be a clear signal for new physics and help to understand the source of neutrino masses and CP violation.
We propose a novel experiment searching for the decay $\mu \rightarrow eee$ with the aim of ultimately reaching a sensitivity of $10^{-16}$, an improvement by four orders of magnitude compared to previous experiments. The technologies enabling this step are thin high-voltage monolithic active pixel sensors for precise tracking at high rates with a minimum of material and scintillating fibres for high resolution time measurements.
\end{abstract}
\begin{center}
\small
\emph{
contribution to NUFACT 11,\\ XIIIth International Workshop on Neutrino Factories, Super beams and Beta beams, 1-6 August 2011, CERN and University of Geneva\\
(Submitted to IOP conference series)}
\end{center}

\section{Introduction}

In the \emph{standard model} (SM) of elementary particle physics, the total number of leptons of each flavour is conserved. The discovery of neutrino oscillations \cite{Fukuda:1998mi, Ahmad:2001an, Eguchi:2002dm} revealed that this is a broken symmetry and leptons can change their flavour. \emph{Lepton flavour violation} (LFV) has however never been observed in the charged lepton sector. The exact mechanism and size of LFV being unknown, its study is of large interest, as it is linked to neutrino mass generation, \emph{CP} violation and new physics beyond the SM.

\begin{figure}
\begin{minipage}[b]{0.48\linewidth}
\centering
	\includegraphics[width=0.45\textwidth]{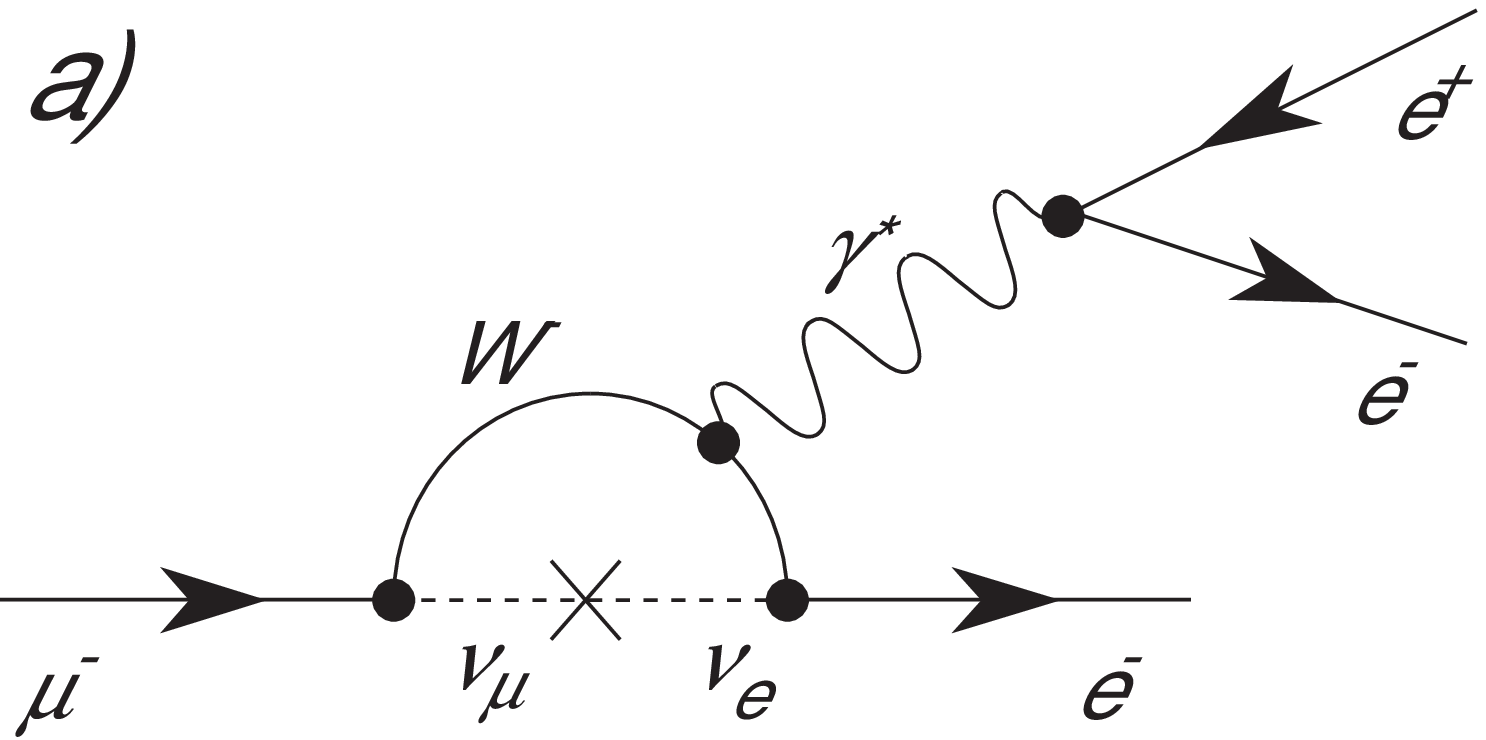}\\ \vspace{3mm}
	\includegraphics[width=0.45\textwidth]{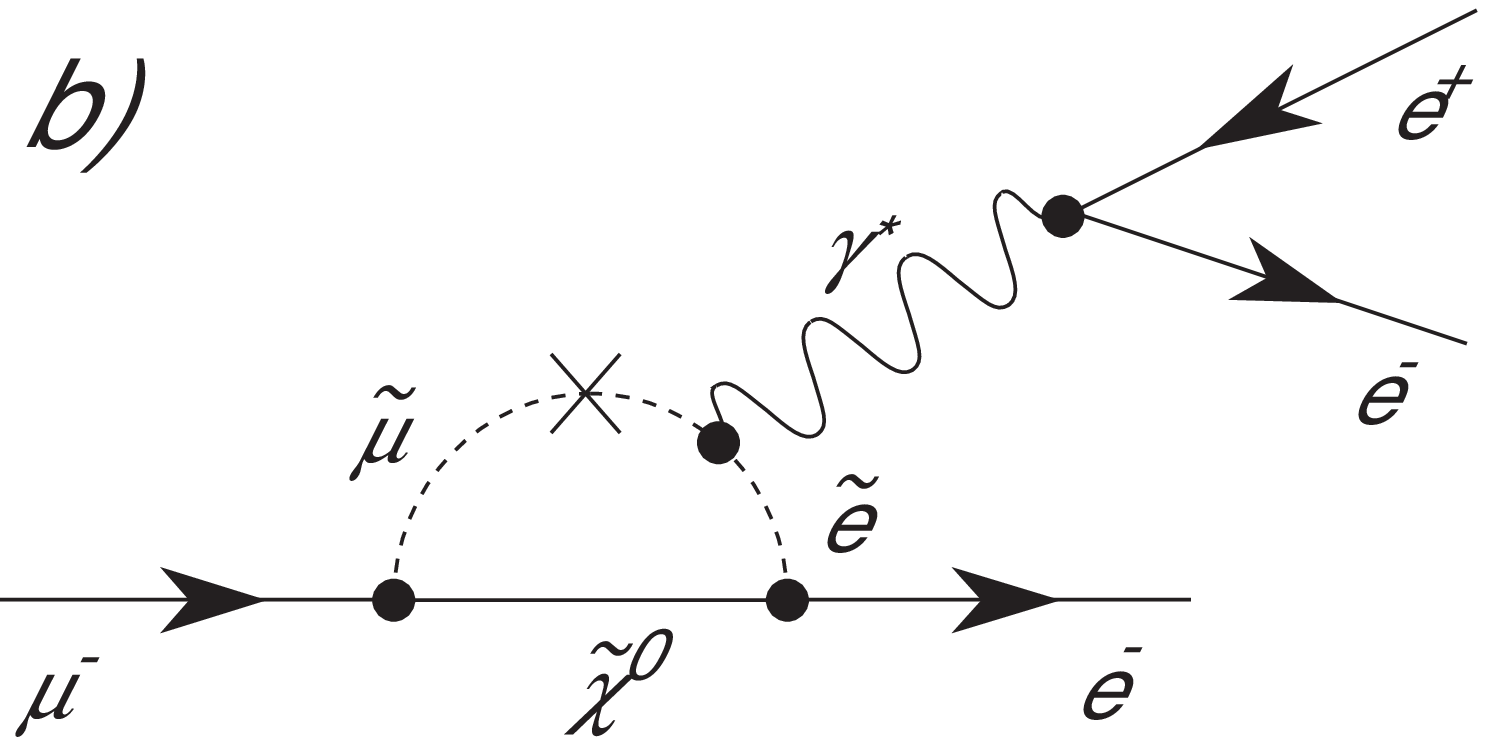}\\ \vspace{3mm}
	\includegraphics[width=0.45\textwidth]{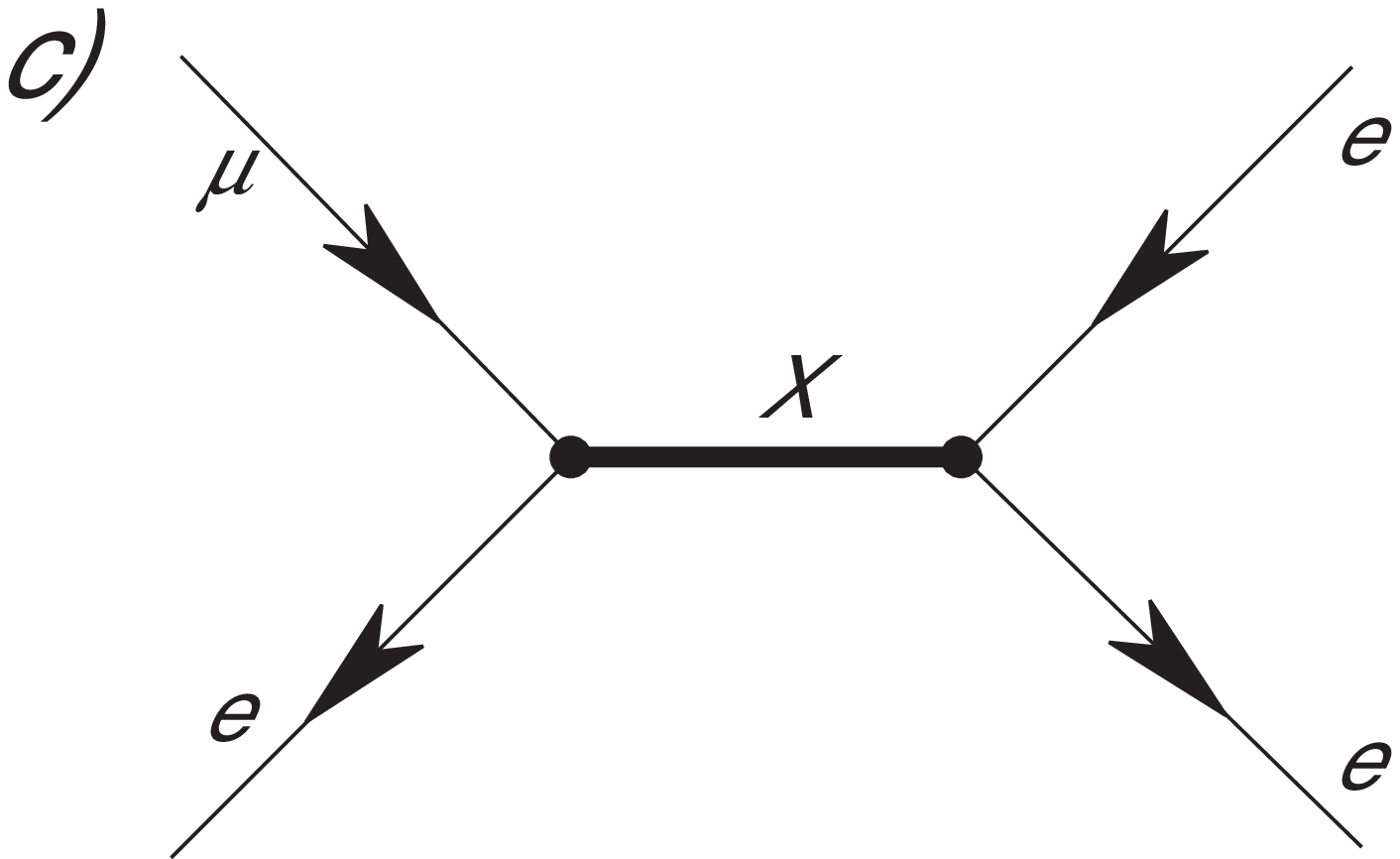}\\ \vspace{3mm}
	\caption{a): Feynman diagram for the \mte process via neutrino oscillation (indicated by the cross). b): Diagram for lepton flavour violation involving supersymmetric particles. c) Diagram for lepton flavour violation at tree level.}
	\label{fig:m3e_neutrino_osc}
\end{minipage}
\hspace{0.02\linewidth}
\begin{minipage}[b]{0.48\linewidth}
\centering
	\includegraphics[width=0.8\textwidth]{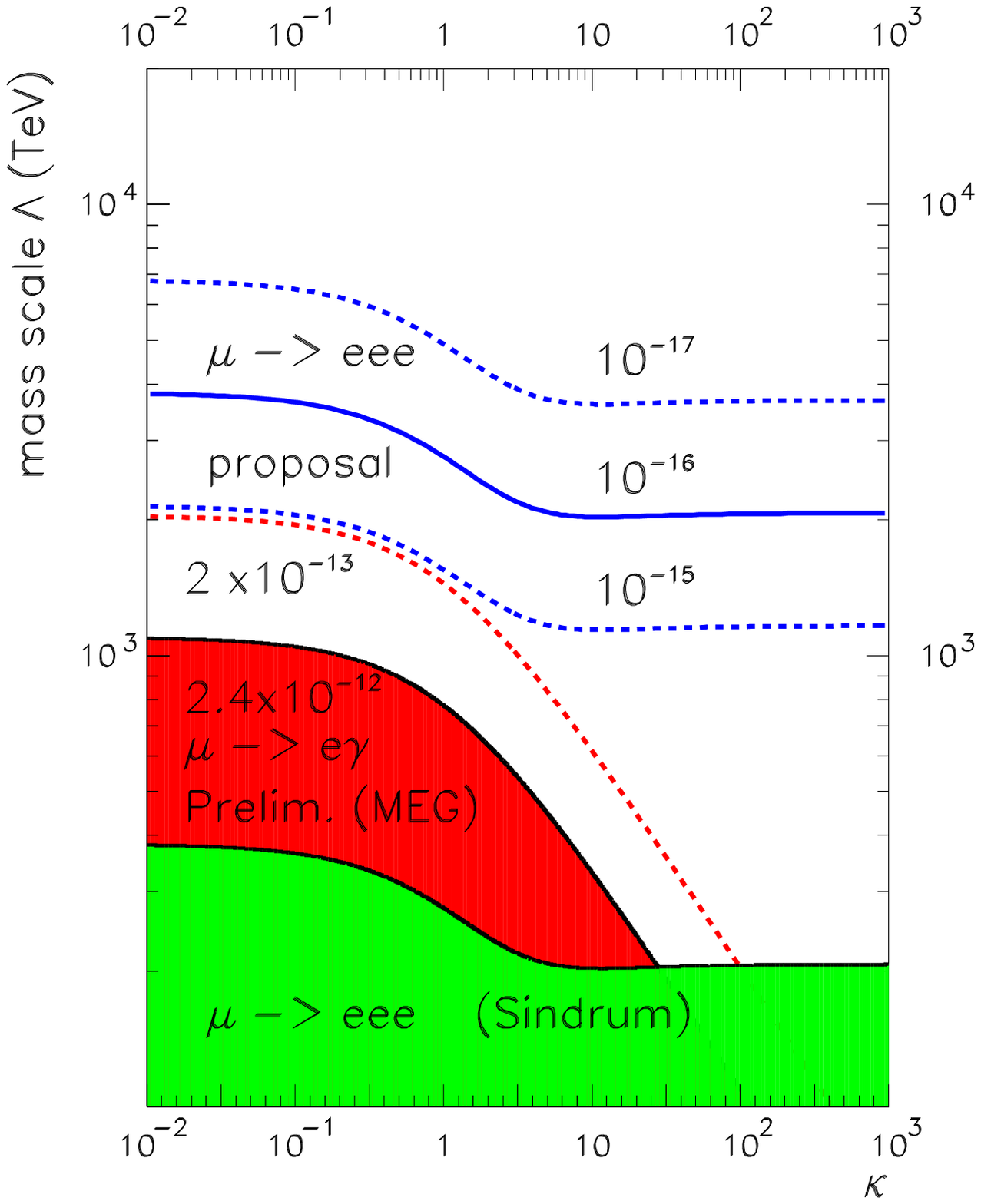}
	\caption{Experimental limits and projected limits on the LFV mass scale $\Lambda$ as a function of the parameter $\kappa$ (see equation \ref{eq:kappa}), taken from \cite{Schoning_CHIPP}.}
	\label{fig:mu3ekappa}
\end{minipage}
\end{figure}


Charged lepton flavour violating reactions can be induced by neutrino mixing in a loop diagram (see figure \ref{fig:m3e_neutrino_osc} a) which is however strongly suppressed, leading to (unobservable) LFV branching fractions in muon decays of less than $10^{-50}$. In many extensions of the SM, such as grand unified models \cite{Pati:1974yy,Georgi:1974sy,Langacker1981}, supersymmetric models \cite{Haber:1984rc} (see figure \ref{fig:m3e_neutrino_osc} b), left-right symmetric models \cite{Mohapatra:1974hk, Mohapatra:1974gc,Senjanovic:1975rk}, models with an extended Higgs sector \cite{Kakizaki:2003jk} and models where electroweak symmetry is broken dynamically \cite{Hill:2002ap}, an experimentally accessible amount of LFV is predicted. The observation of LFV in the charged lepton sector would be a sign for new physics, possibly at scales far beyond the reach of direct observation at e.g.~the large hadron collider (LHC).

Many experiments have been performed or are in operation to search for LFV in the decays of muons or taus. Most prominent are the search for the radiative muon decay \meg \cite{Brooks:1999pu,Nicolo:2003at,Adam:2009ci,Adam:2011ch}, the decay \mte \cite{Bellgardt:1987du}, the conversion of captured muons to electrons \cite{Kaulard:1998rb} and LFV tau decays \cite{HAYASAKA2011}. 

\section{Theory of \mte}

The lepton flavour violating three electron decay of the muon can be mediated either via loops (see figure \ref{fig:m3e_neutrino_osc} a) and b)) or at tree level, via the exchange of a (heavy) new particle (see figure \ref{fig:m3e_neutrino_osc} c)). The effective couplings of the loop diagram are of tensor type (dipole couplings) if the photon is real and receive additional vector-type contributions if the photon is virtual. The four fermion coupling of the tree diagram c) can be described as a contact interaction due to the large mass of the intermediate particle. In order to allow comparisons between \meg and \mte experiments, the following simplified Lagrangian, introducing a common mass scale $\Lambda$ is used:
\begin{equation}
\label{eq:kappa}
	\mathcal{L}_{LFV} = \frac{m_\mu}{(\kappa +1) \Lambda^2}\bar{\mu_R}\sigma^{\mu\nu}e_L F_{\mu\nu} + \frac{\kappa}{(\kappa +1) \Lambda^2} (\bar{\mu_L}\gamma^\mu e_L)(\bar{e_L}\gamma_\mu e_L),
\end{equation}
here the left-left vector coupling is chosen as exemplary for the contact term (right), the parameter $\kappa$ describes the strength of the contact interaction amplitude relative to the loop amplitude. Figure \ref{fig:mu3ekappa} shows limits on the mass scale $\Lambda$ as a function of $\kappa$ for the MEG and SINDRUM experiments, as well as limits expected from the MEG experiment at design sensitivity and for a \mte experiment with sensitivities from $10^{-15}$ to $10^{-17}$. At the same branching fraction, \meg experiments are more sensitive at low values of $\kappa$ (dominance by the loop diagram), whilst \mte experiments constrain $\Lambda$ at high values of $\kappa$. 
In the case of a dominant on-shell photon contribution ($\kappa \rightarrow 0$), a quasi model-independent relation between the \meg and \mte decay rates can be derived:
\begin{equation}
	\frac{B(\mu \rightarrow eee)}{B(\mu \rightarrow e\gamma)} \approx 0.006.
\end{equation}
In order to set competitive constraints on LFV dipole couplings, a limit on the branching fraction of the decay \mte thus has to be about two orders of magnitude smaller than the best \meg limit. In case of a discovery, an \mte experiment (with a three-body final state) would give access to CP observables and thus allow for a search for CP violation. In \meg experiments this would require a (difficult) measurement of the photon polarization.


\section{Challenges for a \mte experiment}

A \mte experiment aiming for a sensitivity of $10^{-16}$ has to be able to suppress accidental backgrounds and the process $\mu \rightarrow eee \nu \bar{\nu}$ (\emph{internal conversion}) at least at the same level whilst running at a rate in excess of $10^9$ muons on target per second, rates which are achievable with a beam-line upgrade at the Paul Scherrer Institut (PSI) in Switzerland. Accidental background can be controlled by an excellent vertex and timing resolution, whilst the only distinguishing feature of the internal conversion decay is the missing momentum carried away by the neutrinos; thus a very precise track reconstruction is needed. 

Conventional tracking detectors either have a limited rate capacity and resolution (gaseous detectors) or induce unacceptable amounts of multiple scattering (hybrid silicon pixel sensors). The development of thin, high-voltage active silicon pixel sensors (HV-MAPS) and scintillating fibre trackers opens up new possibilities which we plan to exploit for a novel experiment searching for the LFV decay \mte.

\section{Detector technologies}

Recently, thin active silicon pixel detectors have become available \cite{Turchetta:2001dy, Peric:2007zz, Peric2010504,Peric2010, Winter2010192, DeMasi2011296}, opening up new possibilities for low-material, high-precision tracking detectors. Produced in a CMOS process, these sensors include amplification and read-out logic inside of pixels with pitches from $20$ to $100~\mu\textrm{m}$. The silicon can be thinned to a thickness of $50~\mu\textrm{m}$ without affecting the signal collection efficiency. For the application in the \mte experiment, the high-voltage CMOS monolithic active pixel sensor (MAPS) technology developed at the Institute for Computer Science of the University of Heidelberg (\emph{ZITI}) is particularly interesting \cite{Peric:2007zz, Peric2010504,Peric2010}. Here a process originally developed for driving automotive and industrial devices with high voltage signals is used to create a monolithic pixel sensor, where all the transistors are placed in a deep N-well, which can be reversely biased with regard to the P-substrate with a high voltage, typically more than $50~\textrm{V}$. This creates a relatively large depleted area at the transition. Particles passing through this depleted zone will generate a sizeable amount of charge (around $2200~e$ for a depletion zone depth of $9~\mu\textrm{m}$, leading to a signal-to-noise ratio above $30$), which is separated by the strong electric field, leading to a fast current signal. The technology is well suited for tracking low momentum particles: because the signal is collected close to the chip surface, the chip can be thinned without signal loss. It is also resistant to both non-ionizing and (with a suitable design of the electronics) ionizing radiation. As a standard industrial process is used, the sensors are relatively cheap. The process has reticles of up to $2 \times 2~\textrm{cm}$ size, which can be combined by wire-bonding or stitching to $2 \times 6~\textrm{cm}$ sensors. 

Without active cooling, the sensors can be read out at 10 to 40~MHz, leading to up to hundred electron tracks per readout frame. Precise timing information helping to suppress accidental background should come from a scintillating fibre tracker with a time resolution better than 100~ps. 

\section{Detector concept}

\begin{figure}
	\centering
		\includegraphics[width=1.00\textwidth]{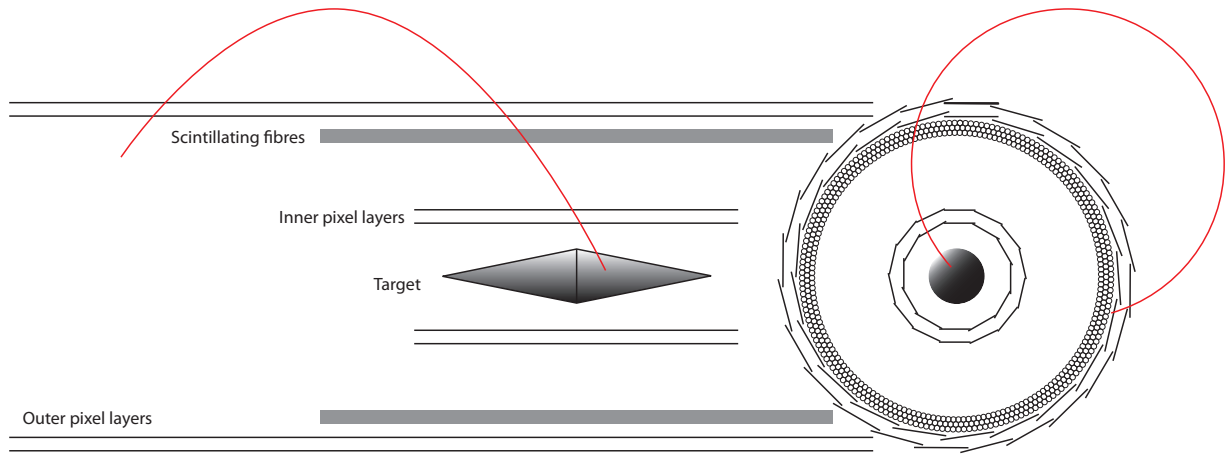}
	\caption{Schematic view of a possible \mte experiment.}
	\label{fig:Schematic-2}
\end{figure}

The best geometry for a detector providing high resolution tracking for momenta from 15 to 53~MeV/$c$ with a large angular coverage is currently under study. Both for resolution in a multiple scattering environment and ease of reconstruction, a spectrometer like arrangement of two double layers separated by free space in a solenoidal magnetic field is preferred. In order to improve the momentum resolution at high momenta, tracks curling back into the detector are reconstructed and fitted. Various detector concepts are currently being studied using a Geant4 \cite{Agostinelli2003250} based simulation. Depending on the exact geometry, the detector will produce a data rate of about 50~Gbit/s, which is more than can practically be stored, but low enough for a triggerless operation mode with a farm of GPU assisted PCs performing fast track fits.

\section{Conclusion}

New detector technologies (especially HV-MAPS), open the opportunity to build an experiment searching for the LFV decay \mte with a sensitivity of up to $10^{-16}$, and thus the potential of finding physics beyond the standard model or strongly constraining models. A collaboration is currently forming and a letter of intent will be submitted to PSI soon.

\section*{References}
\bibliographystyle{iopart-num}
\bibliography{mu3e}

\end{document}